# Dark Energy Constraints on Masses and Sizes of Large Scale Cosmic Structures

*Louise Rebecca*

Christ Junior College, Bangalore - 560 029, India

Telephone: +91-80-4012 9292; Fax: +91-80- 4012 9222

e-mail: louise.rheanna@cjc.christcollege.edu

*Kenath Arun*

Christ Junior College, Bangalore - 560 029, India

Telephone: +91-80-4012 9292; Fax: +91-80- 4012 9222

e-mail: kenath.arun@cjc.christcollege.edu

*C Sivaram*

Indian Institute of Astrophysics, Bangalore - 560 034, India

Telephone: +91-80-2553 0672; Fax: +91-80-2553 4043

e-mail: sivaram@iiap.res.in

**Abstract:** The requirement that their gravitational binding self-energy density must at least equal the background repulsive dark energy density for large scale cosmic structures implies a mass-radius relation of $M/R^2 \approx 1 \, g/cm^2$, as pointed out earlier. This relation seems to hold true for primeval galaxies as well as those at present epoch. This could set constraints on the nature and evolution of dark energy. Besides, we also set constraints on the size of galaxy clusters and superclusters due to the repulsive cosmological dark energy. This could indicate as to why large scale cosmic structures much larger than ~200Mpc are not seen.

In recent works (Sivaram *et al*, 2013; Sivaram and Arun, 2012; 2013) a new type of cosmological paradigm was invoked wherein for a whole hierarchy of large scale structures such as galaxies, galaxy clusters, superclusters, etc. the requirement that their gravitational binding self-energy density (holding them together) must at least equal (or balance) if not exceed the background repulsive dark energy density (assumed to be given by a cosmological constant term Λ, as many current observation would suggest) implies a mass-radius relation of the type:

$$\frac{M}{R^2} = \frac{c^2}{G}\sqrt{\Lambda} \qquad \ldots (1)$$

where $M$ is the total observed mass and $R$ is the observed size of the large scale structures. Equation (1) can be obtained by the balance between the gravitational energy density, $\frac{GM^2}{8\pi R^4}$ and the repulsive dark energy density (caused by a Λ term) of $\frac{\Lambda c^4}{8\pi G}$. Thus:

$$\frac{GM^2}{8\pi R^4} = \frac{\Lambda c^4}{8\pi G} \qquad \ldots (2)$$

Equation (1) follows from equation (2).

If the dark energy is given by a cosmological constant, of the observed value, $\Lambda \approx 10^{-56} cm^{-2}$ (Ade *et al*, 2014), we then have:

$$\frac{M}{R^2} = \frac{c^2}{G}\sqrt{\Lambda} \approx 1 \, g/cm^2 \qquad \ldots (3)$$

As pointed out earlier (Sivaram, 1994; 2000; Sivaram and Arun, 2013; Sivaram *et al*, 2013), this relation (given by equations (2) and (3)) holds from globular clusters to galaxies as well as galaxy clusters up to the Hubble volume. Moreover, the above relations for a wide range of structures was sought to be understood in terms of an extended holographic principle, which was initially proposed for black holes (Sivaram and Arun, 2013).

As shown in the earlier work (Sivaram and Arun, 2013; Sivaram *et al*, 2013), equation (3) hold for all the galaxies observed at low redshifts. For example, in the case of the Andromeda galaxy, we have, $M/R^2 \sim 0.5 \, g/cm^2$; for Messier 77, this ratio is $\sim 0.3 \, g/cm^2$.

In this brief note, we extend this argument to include galaxies at high redshifts that are recently discovered (Oesch *et al*, 2016; Law *et al*, 2012). We show that the above relation given by equations (1) to (3) also holds for primeval galaxies at early epochs. For the oldest galaxy currently known, GN-z11, at a redshift of z = 11.09, this ratio works out to be $\sim 0.22 \, g/cm^2$; and for Q2343-BX442 (at a redshift of z = 2.18) we have $M/R^2 \sim 0.3 \, g/cm^2$. The ratio $(M/R^2)$ for various galaxies, from the early primeval galaxy, to low z galaxies (in the present epoch) are in agreement with the constraint set by equations (1) to (3).

Thus it is remarkable that the 'universal' $M/R^2$ relation implied by equations (1) to (3), with a dark energy background cosmological constant also holds for galaxies both in the early universe and at the present epoch. This in turn is suggestive of the evolution of dark energy, i.e. it appears that it has remained constant from the early formation of galaxies. Here we suggest that this scenario is consistent with a constant cosmological constant as the background dark energy density.

It is also of interest to note that as far as large galaxy clusters and superclusters are concerned (like Coma Cluster or the Laniakea Supercluster) the repulsive cosmological dark energy can limit the sizes to which such clusters can grow, i.e., to the point where the attractive gravitational force matches the repulsive dark energy density. For a gravitationally bound structure, we have:

$$\frac{GM}{R} - \frac{1}{3}\Lambda c^2 R^2 \geq 0 \qquad \ldots (4)$$

Where, $\frac{GM}{R}$ is the gravitational potential and $\frac{1}{3}\Lambda c^2 R^2$ is the potential due to dark energy. Equation (4) implies constraints on the sizes of large gravitationally bound structures such as galaxy clusters and superclusters. Their size is limited as:

$$R_{lim} \leq \left(\frac{3GM}{\Lambda c^2}\right)^{1/3} \qquad \ldots (5)$$

The size constraint from equation (5), closely matches with the observed sizes of clusters and superclusters, as can be seem from table 1.

| Large scale structure | $R_{lim}$ (in cm) | $R_{obs}$ (in cm) |
|---|---|---|
| Virgo Cluster | $3.7 \times 10^{25}$ | $7.09 \times 10^{24}$ |
| Coma Cluster | $3.13 \times 10^{25}$ | $9.46 \times 10^{24}$ |
| Phoenix Cluster | $3.4 \times 10^{25}$ | $1 \times 10^{24}$ |
| Saraswati Supercluster | $1 \times 10^{26}$ | $2 \times 10^{26}$ |
| Laniakea Supercluster | $1.63 \times 10^{26}$ | $2.36 \times 10^{26}$ |
| Horologium Supercluster | $1.61 \times 10^{26}$ | $5 \times 10^{26}$ |
| Corona Borealis Supercluster | $7.47 \times 10^{25}$ | $3.1 \times 10^{26}$ |

Table 1: Observed and calculated sizes of clusters and superclusters

This may explain why no structures larger than about 200Mpc is seen.

In short, the universality of the $M/R^2$ relation can be suggestive of the evolution of dark energy. It appears that dark energy has remained constant from the early formation of galaxies, indicating a constant cosmological constant as the background dark energy density. Moreover, the repulsive cosmological dark energy can limit the sizes to which galaxy clusters and superclusters can grow. Their observed sizes are close to this limit (as seen from Table 1), which hint as to why large scale cosmic structures larger than what are observed cannot form.

In the above discussion we have addressed the implication of dark energy (as given by cosmological constant) for the limitations on masses and sizes of large scale structures. However, we can stress that dark energy could also arise through extended gravity models as emphasised in (Corda, 2009), which will lead to similar consequences. Also for a review, and discussions on alternate models to dark energy (and dark matter) see (Arun *et al*, 2017; 2018).